# Mediation Analysis for Count and Zero-Inflated Count Data Without Sequential Ignorability and Its Application in Dental Studies

Zijian Guo, Dylan S. Small

*Department of Statistics, University of Pennsylvania Wharton School, Philadelphia, PA 19104, USA*

Stuart A. Gansky, Jing Cheng

*Division of Oral Epidemiology & Dental Public Health, University of California, San Francisco, CA 94143, USA*

E-mail: zijguo@wharton.upenn.edu, dsmall@wharton.upenn.edu, stuart.gansky@ucsf.edu, jing.cheng@ucsf.edu

**Summary**. Mediation analysis seeks to understand the mechanism by which a treatment affects an outcome. Count or zero-inflated count outcome are common in many studies in which mediation analysis is of interest. For example, in dental studies, outcomes such as decayed, missing and filled teeth are typically zero inflated. Existing mediation analysis approaches for count data often assume sequential ignorability of the mediator. This is often not plausible because the mediator is not randomized so that there are unmeasured confounders associated with the mediator and the outcome. In this paper, we develop causal methods based on instrumental variable (IV) approaches for mediation analysis for count data possibly with a lot of zeros that do not require the assumption of sequential ignorability. We first define the direct and indirect effect ratios for those data, and then propose estimating equations and use empirical likelihood to estimate the direct and indirect effects consistently. A sensitivity analysis is proposed for violations of the IV exclusion restriction assumption. Simulation studies demonstrate that our method works well for different types of outcomes under different settings. Our method is applied to a randomized dental caries prevention trial and a study of the effect of a massive flood in Bangladesh on children's diarrhea.

## 1. Introduction

In many studies, the intervention is designed to change intermediate variables under the hypothesis that the change in those intermediate variables will lead to improvement in the final outcomes (MacKinnon and Luecen,2011). In these studies, in addition to the overall effect of the intervention on the outcome in the end, researchers would like to know whether and how much the intervention affects the outcome through the measured intermediate variables (mediators) as designed (indirect effect) vs. "direct" intervention effects on the outcome not through the proposed mediators but involving other unknown pathways. Knowing those effects helps us to understand the working mechanism of an intervention and to tailor specific intervention components for future research and applications in specific populations.

Standard mediation approaches since Baron and Kenny (Baron and Kenny, 1986; Cole and Maxwell 2003; MacKinnon 2008), such as regression, path and structural equation model (SEM) among others, assume sequential ignorability of the intervention and mediator (effective randomization of the intervention and mediators) for a causal interpretation on the direct and mediation effects of the intervention on the outcome. Although the ignorability assumption for the intervention is reasonable when the intervention is randomized, the ignorability assumption for mediators can be questionable because mediators are not randomized by researchers so that there may be unmeasured confounders between the mediators and the outcome. Recently developed causal methods (Robins and Greenland 1992; Pearl 2001; Rubin 2004; Ten Have et al. 2007; Albert, 2008; van der Laan and Petersen 2008; Sobel 2008; Goetgeluk et al. 2009; VanderWeele and Vansteelandt 2009, 2010; Elliott et al. 2010; Imai et al. 2010; Jo et al. 2011; Daniels et al. 2012; Stayer et al. 2014) adopt the potential outcome framework and make different assumptions about the intervention and mediator to estimate the indirect (mediation) and direct effect of the intervention on the outcome. Some of these causal methods have replaced the ignorability assumption of mediators with other assumptions, such as the no interaction assumption between intervention and mediator among others.

Most standard and causal approaches focus on continuous and/or binary mediators and outcomes. However, the outcome variable in many studies is often a count following a Poisson or Negative Binomial distribution, or a zero-inflated count that has a higher probability of being zero than expected under a Poisson or Negative Binomial. Zero inflated count outcomes are particularly common in caries studies, where the primary outcome of interest is often the number of decayed, missing and filled teeth (dmft/DMFT) or surface indices (dmfs/DMFS) a subject has. The development of caries is a long time process during which pathological factors and protective factors work against one another (Featherstone 2003), so in a population with a large proportion of low caries risk subjects or young children with a relatively short time exposure to pathological factors, most subjects do not have any caries and therefore dmft/DMFT and dmfs/DMFS are counts with a lot of zeros (Ismail et al. 2011, Featherstone et al. 2012). Zero inflated count outcomes are also common in other settings such as number of days a child is sick from a cause like diarrhea or respiratory illness, number of healthcare visits and number of days stayed in a hospital (Cameron and Trivedi, 1998).

Assuming sequential ignorability, Albert and Nelson (2011) discussed a nice generalized mediation approach with application to count dental outcomes, Wang and Albert (2012) provided a mediation formula for the me-diation effect estimation in a two-stage model and considered a decomposition of the mediation effect in a three-stage model with application to zero-inflated count outcomes, and Albert (2012) considered an inverse-probability weighted estimator for the mediation effect on count outcomes. Sequential ignorability is plausible for some studies. However, for the study that motivated our work, the Detroit Dental Health Project's Motivational Interviewing DVD (DDHP MI-DVD) study, the sequential ignorability may not be plausible. The DDHP MI-DVD study was a randomized dental trial of a Motivational Interviewing (MI) intervention to prevent early childhood caries (ECC) in low income African-American children ($0-5$ years) in Detroit, Michigan (Ismail et al. 2011). In the study, caregivers in both intervention and control groups watched a 15-minute education video on children's oral health. For the intervention group, a MI interviewer reviewed the child's dental examination with caregivers, and discussed caregivers' personal thoughts and concerns about specific goals for their child's oral health. A brochure with caregivers' specific goals and a general list of 10 recommendations on diet, oral hygiene and dental visits was given to caregivers in the intervention and control groups, respectively. The outcome of interest is the number of new cavitated, and new untreated lesions 2 years later (Ismail, 2011). Since the majority of the children did not have any new cavitated and new untreated lesions yet at the end of the study, the distribution of the outcome contains a lot of zeros (Figure 1(a)). We are interested in whether and how much the MI intervention prevented new cavities in children through its effect in changing parents/children's oral health behaviors (e.g. children brushed teeth or parents made sure their children brush teeth). Parents/children's oral health behaviors were not randomized or controlled by the investigators and could be affected by factors other than the MI intervention, such as oral health education offered by primary dentists, at school or community, or via internet, which is an unmeasured confounder between parents/children's oral health behavior and their final dental outcomes. Consequently, sequential ignorability may not be plausible in this study.

Therefore, in this paper, we develop a new mediation analysis based on the instrumental variable (IV) approach for count and zero-inflated count data, when there is a concern about unmeasured confounding such that the assumption of sequential ignorability might fail. Our new approach does not require a parametric distribution assumption and sequential ignorablity assumption. When confounding is of concern in a study, IV methods are very helpful for obtaining accurate estimates for treatment effects by adjusting for both unmeasured and measured confounders when a valid IV can be found (Angrist et al., 1996). Angrist and Krueger (1991) provides a good review of applications of the IV method. Methods for mediation analysis based on the IV approach have been proposed by investigators (Ten Have et al., 2007; Albert, 2008; Dunn and Bentall, 2007; Small, 2012) using the randomization interacted with baseline covariates as IVs, but those methods focus on linear models for continuous outcomes. When the outcome model is linear, two stage least squares (2SLS) and two stage residual inclusion (2SRI) can estimate the direct and indirect effects well when there is a valid IV. We will show that the two stage method can give a biased estimate when the mediator is binary and the outcome model is a count or zero-inflated count model. We will first define the direct and indirect treatment effects in our context and then develop a consistent estimator based on estimating equations and empirical likelihood. We will use the random assignment interacted with baseline covariates as IVs to account for both measured and unmeasured confounding. Since the randomized treatment itself is not used as the IV, we are able to estimate the direct and indirect effect of the treatment on the outcome of interest. We also develop a novel method of partial testing the assumption that random assignment interacted with baseline covariates are valid IVs. Although we focus on count and zero-inflated count outcomes in this paper, the method can be generalized to other types of outcomes.

The paper is organized as follows. In Section 2, we introduce notation, framework, and the direct and indirect treatment effects of interest for count and zero-inflated count data. In Section 3, we introduce the IV and two stage approaches, and our new method, and provide a sensitivity analysis method. We present simulation studies in Section 4, and in Section 5, apply our method to the Detroit Dental study and another example, a study of the effective of a massive flood in Bangladesh on childrens diarrhea. In Section 6, future research directions are discussed. The proofs and further simulation studies are provided in the supplementary materials.

## 2. Notation, Framework and Causal Effects of Interest

*Notation* We adopt the potential (counterfactual) outcome framework (Neyman, 1923; Rubin, 1974) and use $Z_i = z$ ($z = 0$ or 1) for the randomly assigned treatment for subject $i$; let $M_i^z$ denote the potential value of a mediator under treatment $z$; use $Y_i(z, m)$ to denote the potential outcome subject $i$ would have under the treatment $z$ and mediator $m$, and $Y_i(z, M_i^z)$ for the potential outcome subject $i$ would have under $Z_i = z$ (where $M_i^z$ would be at its "natural" level under $z$). We let $U_i$ denote unobserved confounders, $\mathbf{X}_i$ denote observed baseline covariates, and $\mathbf{X}_i^{\text{IV}}$ denote a subset of baseline covariates to construct IVs.

*Direct and Indirect Effects of Interest* For count and zero-inflated count outcomes, we are particularly interested in the controlled and natural direct and indirect ratios for comparing average potential outcomes at different levels of randomization and mediator.

$$\text{Controlled effect ratio:}\quad \text{direct}\,(z \text{ vs. } z^*; m, \mathbf{x}, u) \quad \frac{\mathbb{E}\left(Y\left(z,m\right)|\mathbf{x},u\right)}{\mathbb{E}\left(Y\left(z^*,m\right)|\mathbf{x},u\right)}, \tag{1}$$

$$\text{indirect}\,(m \text{ vs. } m^*; z, \mathbf{x}, u) \quad \frac{\mathbb{E}\left(Y\left(z,m\right)|\mathbf{x},u\right)}{\mathbb{E}\left(Y\left(z,m^*\right)|\mathbf{x},u\right)}; \tag{2}$$

$$\text{Natural effect ratio:}\quad \text{direct}\left(z \text{ vs. } z^*; M^{z^*}, \mathbf{x}, u\right) \quad \frac{\mathbb{E}\left(Y\left(z,M^{z^*}\right)|\mathbf{x},u\right)}{\mathbb{E}\left(Y\left(z^*,M^{z^*}\right)|\mathbf{x},u\right)}, \tag{3}$$

$$\text{indirect}\left(M^{z} \text{ vs. } M^{z^*}; z, \mathbf{x}, u\right) \quad \frac{\mathbb{E}\left(Y\left(z,M^{z}\right)|\mathbf{x},u\right)}{\mathbb{E}\left(Y\left(z,M^{z^*}\right)|\mathbf{x},u\right)} \tag{4}$$

A ratio of 1 indicates no effect. The controlled direct effect sets the mediator at a fixed value ($m$) and the natural direct effect sets the mediator at its "natural" level that would be achieved under treatment assignment $z^*$ ($M^{z^*}$). The natural indirect effect ratio tells us the ratio of average outcomes under treatment $z$ that would be observed if the mediator would change from the value under a treatment $z$ ($M^z$) to the value under another treatment $z^*$ ($M^{z^*}$). The expectations in (1) and (2) are taken over the conditional distribution of the potential outcome. However, in the natural effect ratio, since the mediator is random, the expectations in (3) and (4) are taken over the conditional joint distribution of the mediator and the potential outcome corresponding to the mediator. We will discuss below the settings in which the controlled and natural effects are identified.

*The Model Setting* We consider the following generalized linear model for the expected potential outcomes:

$$f\{\mathbb{E}\left(Y(z,m)|\mathbf{x},u\right)\} = \beta_0 + \beta_z z + \beta_m m + \beta_{\mathbf{x}}\mathbf{x} + \beta_u u, \tag{5}$$

where $f$ is a link function. The generalization of Model (5) to include the interaction terms $z \times \mathbf{x}$,$z \times m$ and $\mathbf{x} \times m$ will be discussed in Section 6. For a Poisson or Negative Binomialcount outcome, a log link function will be used in the model; For a zero-inflated count outcome, we consider a Neyman

Type A distributed outcome (Dobbie and Welsh, 2001), $y = \sum_{k=1}^{N} y_k$, where

$$N|\mathbf{x}, z, m, u \sim \mathit{Poisson}\left(\exp\left(\gamma_0 + \gamma_z z + \gamma_m m + \gamma_{\mathbf{x}}\mathbf{x} + \gamma_u u\right)\right);$$

$$y_k|\mathbf{x}, z, m, u \sim \mathit{Poisson}\left(\exp\left(\lambda_0 + \lambda_z z + \lambda_m m + \lambda_{\mathbf{x}}\mathbf{x} + \lambda_u u\right)\right).$$

The Neyman Type A distribution is also a special case of (5) with

$$\log \mathbb{E}(Y(z,m)|\mathbf{x}, u) = (\gamma_0+\lambda_0)+(\gamma_z+\lambda_z)z+(\gamma_m+\lambda_m)m+(\gamma_{\mathbf{x}}+\lambda_{\mathbf{x}})\mathbf{x}+(\gamma_u+\lambda_u)u.$$

Without loss of generality, we will assume that $\mathbb{E}\left(\exp\left(\beta_u u\right)\right) = 1$ for the count and zero-inflated count with a log link function.

<u>*Controlled Effect Ratios*</u> Given Model (5) with a log link function, we have:

$$\frac{\mathbb{E}\left(Y\left(z,m\right)|\mathbf{x},u\right)}{\mathbb{E}\left(Y\left(z^*,m\right)|\mathbf{x},u\right)} = \exp\left(\beta_z(z-z^*)\right), \quad \frac{\mathbb{E}\left(Y\left(z,m\right)|\mathbf{x},u\right)}{\mathbb{E}\left(Y\left(z,m^*\right)|\mathbf{x},u\right)} = \exp\left(\beta_m(m-m^*)\right). \quad (6)$$

Therefore, estimating the controlled effect ratios is equivalent to estimating $\beta_z$ and $\beta_m$.

<u>*Natural Effect Ratios*</u> For natural effect ratios, Model (5) becomes

$$f\{\mathbb{E}\left(Y(z, M^{z^*})|\mathbf{x}, u\right)\} = \beta_0 + \beta_z z + \beta_m M^{z^*} + \beta_{\mathbf{x}}\mathbf{x} + \beta_u u, \quad (7)$$

and we further consider a model for the mediator:

$$h\left(\mathbb{E}\left(M^{z^*}|\mathbf{x}, u\right)\right) = \alpha_0 + \alpha_z z^* + \alpha_{\mathbf{x}}\mathbf{x} + \alpha_{IV} z^* \mathbf{x}^{\mathrm{IV}} + \alpha_u u, \quad (8)$$

where $h$ is a link function, e.g., identity and logit functions for continuous and binary mediators respectively. Then the natural indirect effect ratio with a continuous mediator will be

$$\frac{\mathbb{E}\left(Y\left(z, M^{z}\right)|\mathbf{x},u\right)}{\mathbb{E}\left(Y\left(z, M^{z^*}\right)|\mathbf{x},u\right)} = \exp\left(\beta_m \alpha_z\left(z-z^*\right) + \beta_m \alpha_{IV}\mathbf{x}^{IV}(z-z^*)\right), \quad (9)$$

and with a binary mediator:

$$\frac{\mathbb{E}\left(Y\left(z, M^{z}\right)|\mathbf{x}, u\right)}{\mathbb{E}\left(Y\left(z, M^{z^{*}}\right)|\mathbf{x}, u\right)} = \frac{P(M^{z}=1|\mathbf{x},u)\exp(\beta_{m}) + P(M^{z}=0|\mathbf{x},u)}{P(M^{z^{*}}=1|\mathbf{x},u)\exp(\beta_{m}) + P(M^{z^{*}}=0|\mathbf{x},u)} \quad (10)$$

The natural direct effect ratio for both continuous and binary mediators will be the same as the controlled direct effect ratio:

$$\frac{\mathbb{E}\left(Y\left(z, M^{z^{*}}\right)|\mathbf{x}, u\right)}{\mathbb{E}\left(Y\left(z^{*}, M^{z^{*}}\right)|\mathbf{x}, u\right)} = \exp\left(\beta_{z}(z - z^{*})\right). \quad (11)$$

The proofs of (9), (10) and (11) are provided in Section 1.1 of the supplementary material. For a continuous mediator, the natural direct and indirect effect ratios (9) and (11) are identifiable given that the parameters $\beta_z$, $\beta_m$, $\alpha_z$ and $\alpha_{IV}$ can be estimated consistently. However, the natural indirect effect ratio for a binary mediator depends on the values of the unmeasured $u$ in (10) and is not identifiable without additional assumptions.

## 3. The Instrumental Variable Approach

When there is a concern about an unmeasured confounder $u$, the instrumental variable (IV) approach is a popular technique for dealing with unmeasured confounding; such unmeasured confounding is not addressed by regular regression and propensity score methods. In the context of mediation analysis, a valid IV is a variable that, given the measured baseline variables: (1) affects the value of the mediator; (2) is independent of the unmeasured confounders; and (3) does not have a direct effect on the outcome other than through its effect on the mediator.

Mediation methods based on the IV approach have been proposed by investigators (Ten Have et al., 2007; Albert, 2008; Dunn and Bentall, 2007; Small, 2012) for continuous outcomes, where baseline covariates interacted with random assignment are used as instrumental variables in linear models. In this study, we will also use baseline covariates interacted with random assignment ($Z \times \mathbf{X}^{\mathrm{IV}}$) as instrumental variables for mediation analysis but in nonlinear models. In the setting of a randomized trial with noncompliance, the randomization $Z$, is often used as an instrument (e.g., Sommer and Zeger, 1991; Greevy et al., 2004 ) under the assumption that the randomization has no direct effect on the outcome. In our setting, when $Z$ is randomized we assume that the randomization is complied with but allow for the randomization itself to have a direct effect. Since the randomization can have a direct effect on the outcome around the mediator (violation of exclusion restriction), we cannot use $Z$ as an IV. Instead, we consider $Z \times \mathbf{X}^{\mathrm{IV}}$ as a possible IV to enable us to estimate both the direct and indirect effects of the treatment on the outcome of interest.

To consistently estimate the direct and indirect (mediation) effects discussed in Section 2 without the commonly used sequential ignorability assumption, we will use $Z \times \mathbf{X}^{\mathrm{IV}}$ as an IV and assume:

(1) The treatment $Z$ is randomized.

(2) The conditional distribution of the unmeasured confounding $P(U|\mathbf{X})$ is the same for all $\mathbf{X}$, which implies that $U$ and $\mathbf{X}$ are independent.

(3) There is an interaction between randomized treatment $Z$ and baseline covariate $\mathbf{X}^{\mathrm{IV}}$ predicting mediator $M$ conditional on $Z$ and $\mathbf{X}$.

(4) The interaction $Z \times \mathbf{X}^{\mathrm{IV}}$ affects the outcome only through its effect on the mediator $M$, conditional on $\mathbf{X}$ and $Z$. This assumption is often called the exclusion restriction assumption and cannot be formally tested. A sensitivity analysis is proposed in Section 3.4 to see how the method behaves when the assumption does not hold.

Assumptions (1) and (2) imply that the instrument $Z \times \mathbf{X}^{\mathrm{IV}}$ is independent of the unmeasured confounder $U$; Assumption (3) states that $Z \times \mathbf{X}^{\mathrm{IV}}$ affects the value of the mediator $M$; Assumption (4) says that $Z \times \mathbf{X}^{\mathrm{IV}}$ does not have a direct effect on the outcome other than through its effect on the mediator.

Hence, under these assumptions, the interaction $Z \times \mathbf{X}^{\mathrm{IV}}$ is a valid IV for the mediation analysis. Note that although the treatment $Z$ and instrumental variable $Z \times \mathbf{X}^{\mathrm{IV}}$ affect the mediator $M$, conditioning on $Z$, $M$ and $\mathbf{X}$, the instrumental variable $Z \times \mathbf{X}^{\mathrm{IV}}$ is assumed to have no direct effect on the outcome $Y$. Similar model assumptions are discussed in Equation (11) in Jo (2002).

### *3.1. Two Stage Approach*

When a linear model is a good fit for a continuous outcome, the two stage least square (2SLS) estimator provides consistent estimates for the parameters of interest when a valid IV is available. When the outcome is not continuous such that a linear model does not fit well, two-stage predictor substitution (2SPS) and two-stage residual inclusion (2SRI) (Nagelkerke et al., 2000; Terza et al., 2008) have been proposed to evaluate the treatment effect with an IV. For a binary treatment and outcome, the 2SPS approach fits a logistic regression of the treatment on the IV and covariates in the first stage; and then in the second stage uses the predicted treatment from the first stage to fit a logistic regression for the outcome with covariates. Because of the use of predicted treatment in the second stage, this approach is called two-stage predictor substitution. The first stage of 2SRI approach is the same as that of 2SPS, but in the second stage, instead of using predicted treatment, 2SRI uses the residual from the first stage regression along with observed treatment and covariates to model the outcome in a logistic regression. Cai et al. (2011) showed that 2SPS and 2SRI estimators are asympotitcally biased for the complier average causal effect (CACE) when there is unmeasured confounding. In the supplementary materials, we compare the 2SRI and 2SPS estimators for mediation analysis with simulation studies. When the second stage linear model is a good fit for the outcome, both 2SPS and 2SRI estimates are consistent (Wooldridge, 2010). When the second stage outcome model is nonlinear, 2SRI is approximately unbiased while 2SPS is slightly biased for the Neyman Type A outcome given a linear first stage model for a continuous mediator (see Table 8 in the supplementary material); but neither 2SPS and 2SRI estimate is consistent given a non-linear first stage model for a binary mediator (see Table 7 in the supplementary material).

Because 2SRI performs better than 2SPS, we will examine and compare the performance of 2SRI to the performance of a new approach discussed in Section 3.2 for the mediation analysis in this paper. With the IV $Z \times \mathbf{X}^{IV}$, the 2SRI fits the models in two stages. Stage I: $h\{\mathbb{E}\left(M|z, \mathbf{x}, z \times \mathbf{x}^{IV}\right)\} = \alpha_0 + \alpha_z z + \alpha_m m + \alpha_{IV} z \times \mathbf{x}^{IV} + \alpha_{\mathbf{x}}\mathbf{x}$, where $h$ is a link function; Stage II, $f\{\mathbb{E}\left(Y|m, z, \mathbf{x}, \hat{r}\right)\} = \beta_0 + \beta_z z + \beta_m m + \beta_{\mathbf{x}}\mathbf{x} + \beta_r \hat{r}$, where $\hat{r}$ is the residual $(m - \hat{m})$ from Stage I and $f$ is a link function such as log link for count data.

To see how 2SRI works, we can decompose $U$ into two parts $U = \tau R + \delta$, where $R$ denotes the population residual from the first stage, $\delta$ is the population residual and $\mathbb{E}\left(\delta|R\right) = 0$. Then for continuous and count outcomes, we respectively have:

$$
\begin{aligned}
\mathbb{E}(Y(Z, M)|\mathbf{X}, R) &= \int \beta_0 + \beta_z Z + \beta_m M + \beta_{\mathbf{x}}\mathbf{X} + \beta_u \tau R + \beta_u \delta dP\left(\delta|Z, M, \mathbf{X}, R\right) \qquad (12)\\
&= \beta_0 + \beta_z Z + \beta_m M + \beta_{\mathbf{x}}\mathbf{X} + \beta_u \tau R + \int \beta_u \delta dP\left(\delta|Z, M, \mathbf{X}, R\right).\\
\mathbb{E}\left(Y(Z, M)|\mathbf{X}, R\right) &= \int \exp\left(\beta_0 + \beta_z Z + \beta_m M + \beta_{\mathbf{x}}\mathbf{X} + \beta_u \tau R + \beta_u \delta\right) dP\left(\delta|Z, M, \mathbf{X}, R\right) \qquad (13)\\
&= \exp\left(\beta_0 + \beta_z Z + \beta_m M + \beta_{\mathbf{x}}\mathbf{X} + \beta_u \tau R\right) \int \exp\left(\beta_u \delta\right) dP\left(\delta|Z, M, \mathbf{X}, R\right).
\end{aligned}
$$

<u>Continuous mediators</u> For a continuous mediator, we consider a linear model:

$$
M = \alpha_0 + \alpha_z Z + \alpha_{\mathbf{x}}\mathbf{X} + \alpha_{IV} Z \times \mathbf{X}^{IV} + \alpha_u U + V, \qquad (14)
$$

where $V$ is random error and $U$ is the unmeasured confounder with $(V, U)$ following bivariate normal distribution and is independent of $(\mathbf{X}, Z, Z \times \mathbf{X}^{IV})$. 2SRI fits a linear model for M on $Z, \mathbf{X}, Z \times \mathbf{X}^{IV}$, and the probability limit $\alpha_j^*$ of first stage estimator is equal to the underlying truth, that is, $\alpha_j^* = \alpha_j$, where $j = 0, z, \mathbf{x}, IV$. Then the residual is

$$R = M - \left(\alpha_0 + \alpha_z Z + \alpha_{\mathbf{x}}\mathbf{X} + \alpha_{IV} Z \times \mathbf{X}^{IV}\right) = \alpha_u U + V.$$

Since $(\alpha_u U + V, U)$ is independent of $(\mathbf{X}, Z, Z \times \mathbf{X}^{IV})$, then $\delta$, as the population level residual of regressing $U$ on $R = \alpha_u U + V$, is independent of $(\mathbf{X}, Z, Z \times \mathbf{X}^{IV})$. Since $\delta$ is independent of $R$ and $M$ is linear combination of $(\mathbf{X}, Z, Z \times \mathbf{X}^{IV})$ and $R$, then $\delta$ is independent of $R$ and $M$, and it is easy to see that the 2SRI estimator is consistent for continuous outcomes with a continuous mediator. For count outcomes, because $\delta$ is independent of other variables, $\int \exp\left(\beta_u \delta\right) dP\left(\delta | Z, M, \mathbf{X}, R\right)$ in (13) is a constant. Therefore, the 2SRI estimator is also consistent for count outcomes when the mediator is continuous.

Binary mediators For a binary mediator, we consider a logit model:

$$M | \mathbf{X}, Z, U \sim Ber\left(\frac{\exp(\alpha_0 + \alpha_z Z + \alpha_{\mathbf{x}}\mathbf{X} + \alpha_{IV} Z \times \mathbf{X}^{IV} + \alpha_u U)}{1 + \exp(\alpha_0 + \alpha_z Z + \alpha_{\mathbf{x}}\mathbf{X} + \alpha_{IV} Z \times \mathbf{X}^{IV} + \alpha_u U)}\right), \quad (15)$$

2SRI fits a logit model for M on $Z, \mathbf{X}, Z \times \mathbf{X}^{IV}$, and the probability limit $\alpha_j^*$ of the first stage estimators is not equal to the underlying truth $\alpha_j$, for $j = 0, z, \mathbf{x}, IV$. Then the population residual is

$$R = M - \frac{\exp\left(\alpha_0^* + \alpha_1^* Z + \alpha_2^* \mathbf{X} + \alpha_3^* Z \times \mathbf{X}^{IV}\right)}{1 + \exp\left(\alpha_0^* + \alpha_1^* Z + \alpha_2^* \mathbf{X} + \alpha_3^* Z \times \mathbf{X}^{IV}\right)}.$$

Now $\int \exp\left(\beta_u \delta\right) dP\left(\delta | Z, M, \mathbf{X}, R\right)$ is generally not a constant but a function depending on $Z, M, \mathbf{X}, R$, so the 2SRI estimate will typically be biased when both the mediator and outcome models are nonliear. Proposition 1 shows that if either stage is a linear model with normal error, 2SRI is a consistent estimator under some regularity conditions. The proof is in the supplementary materials. Simulations in Section 4 show that 2SRI can have a large bias when the mediator is binary for a count outcome or the outcome distribution is misspecified.

PROPOSITION 1. *Under regularity conditions, the 2SRI estimator is consistent for (1) a count outcome* (13) *and normal mediator* (14)*; (2) a normal outcome* (12) *and binary mediator* (15)*; and (3) a normal outcome* (12) *and normal mediator* (14)*.*

### *3.2. Estimating Equations and Empirical Likelihood Approach (EE-EL)*

As discussed above, the 2SRI estimate may not be consistent for a count outcome with binary mediators. In this section, we consider a different approach to consistently estimate the parameters of interest in nonlinear models with unmeasured confounding even when the mediator is binary. Unlike 2SRI, this approach does not need to specify the outcome distribution and hence will be robust to the misspecification of the outcome distribution. We let $g(w, \theta) = (g_1(w, \theta), \cdots, g_r(w, \theta))^\intercal$ be estimating functions such that $E\{g(w, \theta)\} = 0$, where $w = (z, \mathbf{x}, m, y)$ and $\theta = (\beta_0, \beta_z, \beta_m, \beta_{\mathbf{x}})$ are the parameters associated with the outcome model. We consider a set of estimating functions (16) to combine information about the parameters and distribution. Under Assumptions (1)-(4), we have $E\{g(w, \theta)\} = 0$. The proof is provided in Section 1.2 of the supplementary material. Equations in (16) include more estimating equations than parameters and consequently there will not typically be a solution that satisfies all the estimating equations. Qin and Lawless (1994) proposed to use Owens (1988, 1990) empirical likelihood approach when there are more estimating equations than parameters and showed that empirical likelihood provides asymptotically efficient estimates of the parameters (in the sense of Van der Vaart(1988) and Bickel, Klaassen, Ritov and Wellner(1993)) under the semiparametric model given by the estimating equations.

$$
\begin{aligned}
g_1(w,\theta) &= \left(\frac{y}{\exp(\beta_0+\beta_z z+\beta_m m+\beta_{\mathbf{x}}\mathbf{x})}-1\right);\\
g_2(w,\theta) &= \left(\frac{y}{\exp(\beta_0+\beta_z z+\beta_m m+\beta_{\mathbf{x}}\mathbf{x})}-1\right)z;\\
g_3(w,\theta) &= \left(\frac{y}{\exp(\beta_0+\beta_z z+\beta_m m+\beta_{\mathbf{x}}\mathbf{x})}-1\right)\mathbf{x};\\
g_4(w,\theta) &= \left(\frac{y}{\exp(\beta_0+\beta_z z+\beta_m m+\beta_{\mathbf{x}}\mathbf{x})}-1\right)\mathbf{x}z;\\
g_5(w,\theta) &= \left(\frac{y}{\exp(\beta_m m)}-\exp(\beta_0+\beta_z z+\beta_{\mathbf{x}}\mathbf{x})\right);\\
g_6(w,\theta) &= \left(\frac{y}{\exp(\beta_m m)}-\exp(\beta_0+\beta_z z+\beta_{\mathbf{x}}\mathbf{x})\right)z;\\
g_7(w,\theta) &= \left(\frac{y}{\exp(\beta_m m)}-\exp(\beta_0+\beta_z z+\beta_{\mathbf{x}}\mathbf{x})\right)\mathbf{x};\\
g_8(w,\theta) &= \left(\frac{y}{\exp(\beta_m m)}-\exp(\beta_0+\beta_z z+\beta_{\mathbf{x}}\mathbf{x})\right)\mathbf{x}z;
\end{aligned}
\tag{16}
$$

Following their approach, we let $p_i$ be the probability of data $(Z_i, \mathbf{X}_i, M_i, Y_i)$ being observed and maximize $\prod_{i=1}^n p_i$ subject to the restrictions

$$p_i \geq 0, \quad \sum_{i=1}^n p_i = 1, \quad \sum_{i=1}^n p_i g(w_i,\theta) = 0.$$

It is equivalent to minimize

$$l_E(\theta) = \sum_{i=1}^n \log\left(1 + t^\tau(\theta) g(w_i,\theta)\right), \tag{17}$$

where $t = (t_1, ..., t_r)^\intercal$ are Lagrange multipliers and are determined by $\frac{1}{n}\sum_{i=1}^n \frac{g(w_i,\theta)}{1+t^\intercal g(w_i,\theta)} = 0$.

With the first four estimating equations $g_1(w,\theta), ..., g_4(w,\theta)$ for $\theta = (\beta_0, \beta_z, \beta_m, \beta_{\mathbf{x}})^\intercal$, the maximized empirical likelihood estimate (MELE) will be the solution to the estimating equations $\sum_{i=1}^n g_j(w_i,\theta) = 0, j = 1, ..., 4$ that minimize (17).

We carry out the computation of maximizing the empirical likelihood subject to the estimating equations being satisfied in two steps: (1) Fix $\theta$, we will minimize (17) with respect to $t$; and (2) Given $t$ from the first step, minimize (17) with respect to $\theta$. Qin and Lawless(1994) showed that the MELE for the estimating equations is consistent under some regularization conditions. Proposition 2 provides the theory that the EE-EL estimator is consistent under some mild regularity conditions.

PROPOSITION 2. *Assume that* $\mathbb{E}\left(g\left(w,\theta_0\right)g^{\intercal}\left(w,\theta_0\right)\right)$ *is positive definite and the rank of* $\mathbb{E}\left(\frac{\partial g(w,\theta)}{\partial\theta}\right)$ *is the same as the dimension of* $\theta$ *and* $\left\|\frac{\partial^2 g(w,\theta)}{\partial\theta\partial\theta^{\intercal}}\right\|$ *can be bounded by some integrable function* $G\left(w\right)$ *in the neighborhood* $\|\theta-\theta_0\|_2\leq 1$ *of the true value* $\theta_0$*. Let* $\hat{\theta}$ *denote the minimizer of* (17)*, then*

$$\sqrt{n}\left(\hat{\theta}-\theta_0\right)\rightarrow N\left(0,V\right), \quad \text{where } V=\left(\mathbb{E}\left(\frac{\partial g}{\partial\theta}\right)^{\intercal}\mathbb{E}\left(gg^{\intercal}\right)^{-1}\mathbb{E}\left(\frac{\partial g}{\partial\theta}\right)\right)^{-1}.$$

PROOF 1. *It suffices to verify the conditions in Theorem 1 in Qin and Lawless(1994). By the expression of* $g(w,\theta)$*,* $g\left(w,\theta\right)$ *and* $\frac{\partial g(w,\theta)}{\partial\theta}$ *are continuous in a compact neighborhood* $\|\theta-\theta_0\|_2\leq 1$ *of the true value* $\theta_0$*. Hence* $\|g\left(w,\theta\right)\|^3$ *and* $\|\frac{\partial g(w,\theta)}{\partial\theta}\|_2$ *are bounded in this compact neighborhood* $\|\theta-\theta_0\|_2\leq 1$*.* $\frac{\partial^2 g(w,\theta)}{\partial\theta\partial\theta^{\tau}}$ *is continuous in* $\theta$ *in a neighborhood* $\|\theta-\theta_0\|_2\leq 1$ *of the true value* $\theta_0$*.*

### 3.3. Testing the Exclusion Restriction and Sensitivity Analysis

When Assumptions (1)-(4) hold, $Z\times\mathbf{X}^{IV}$ is a valid IV. When the IV $Z\times\mathbf{X}^{IV}$ actually affects the outcome directly, then the exclusion restriction (ER) assumption (Assumption (4)) fails so that the estimators will be biased. The violation of the ER assumption means that the IV could have an effect on the outcome other than through the mediator of interest, a situation where the IV could affect the outcome through other intermediate variables besides the mediator of interest. Thus, although the ER assumption can not be for-mally tested, we can partially test the ER assumption by examining whether the IV has an effect on other known intermediate variables besides the mediator of interest. See the supplementary material Section 1.5 for details and assumptions underlying this partial test of the exclusion restriction.

Unfortunately, this test cannot identify all possible ways in which the exclusion restriction could be violated. Consequently, in the remainder of this section, we propose a sensitivity analysis to allow $Z \times \mathbf{X}^{IV}$ to affect the outcome directly by a specified magnitude and then examine how the results will change. Specifically, we consider

$$g(E(Y^{z,m}|z,m,\mathbf{x},u)) = \beta_0 + \beta_z z + \beta_m m + \beta_{\mathbf{x}}\mathbf{x} + \beta_u u + \eta z \times \mathbf{x}^{IV}, \quad (18)$$

where $\eta$ is the sensitivity parameter for the direct effect of the IV on the outcome. When $\eta = 0$, the ER assumption holds. When $\eta \neq 0$, the ER assumption fails and $Z \times \mathbf{X}^{IV}$ is not a valid IV. Higher values of $|\eta|$ means more severe violation of the ER assumption. For a zero-inflated count following Neyman Type A distribution, we have $Y = \sum_{k=1}^{N} y_k$; where

$$N|\mathbf{x},z,m,u \sim Poisson\left(\exp\left(\gamma_0 + \gamma_z z + \gamma_m m + \gamma_{\mathbf{x}}\mathbf{x} + \gamma_u u + \eta_1 z \times \mathbf{x}^{IV}\right)\right);$$
$$y_k|\mathbf{x},z,m,u \sim Poisson\left(\exp\left(\lambda_0 + \lambda_z z + \lambda_m m + \lambda_{\mathbf{x}}\mathbf{x} + \lambda_u u + \eta_2 z \times \mathbf{x}^{IV}\right)\right),$$

we can represent this model in the form of (18) with $\eta = \eta_1 + \eta_2$. With a log link function in Model (18), we will adjust the outcome as:

$$y^{adj} = \frac{y}{\exp(\eta z \times \mathbf{x}^{IV})},$$

and have a model for the adjusted outcome:

$$\mathbb{E}\left(y^{adj}|\mathbf{x},z,m,u\right) = \exp\left(\beta_0 + \beta_z z + \beta_m m + \beta_{\mathbf{x}}\mathbf{x} + \beta_u u\right).$$

Then we can construct the estimating equations (16) by replacing $y$ with $y^{adj}$ in (16) and (17) and obtain the MELE estimator. We will examine how the sensitivity analysis works in the simulation study.

### 3.4. Multiple Mediators

In addition to settings with one mediator discussed above, in this section, we consider settings with multiple conditionally independent mediators. That is, conditioning on $z, \mathbf{x}, z \times \mathbf{x}^{IV}, u$, the mediators are independent. The outcome model (5) is generalized as

$$g\{\mathbb{E}\left(Y(z,\mathbf{m})|\mathbf{x},u\right)\} = \beta_0 + \beta_z z + \beta_{\mathbf{m}}^{\tau}\mathbf{m} + \beta_{\mathbf{x}}\mathbf{x} + \beta_u u. \tag{19}$$

We need at least the same number of instrumental variables as the number of mediators to construct estimating equations and then use the same approaches discussed above for estimation. We discuss the model and estimating equations for two conditionally independent mediators in the supplementary material, which can be easily generalized to multiple mediators.

## 4. Simulation Study

In this section, we will examine the performance of the methods discussed above in finite samples. We consider outcomes that follow Poisson, Negative Binomial and Neyman Type A distributions with a binary mediator, a normally distributed mediator, and multiple mediators respectively. The randomized treatment $Z$ was generated with $P(Z_i = 1) = 0.5$. We consider one or two (standard normal and binary) covariates, and an unmeasured confounder $U$ with $\mathbb{E}\left(\exp(U)\right) = 1$. Mediators and outcomes were generated based on Models (8) and (5) respectively. In the mediator model (8), $\alpha_{IV}$ represents the strength of the IV $Z \times \mathbf{X}^{IV}$ and $\alpha_u$ represents the strength of endogeneous variable $U$. Table 1 shows the true values of parameters in the outcome models. When there are two mediators, we have two IVs in the models. We consider sample sizes of 500, 1000 and 5000. For each setting, 1000 Monte Carlo replications were performed.

Single Binary Mediator We first examine the performance of each approach with a binary mediator, which is generated as

$$m|\mathbf{x}, z, u \sim Ber\left(\frac{\exp(-0.5 + 0.5z + 0.5x + \alpha_{IV} z \times x^{IV} + 0.5u)}{1 + \exp(-0.5 + 0.5z + 0.5x + \alpha_{IV} z \times x^{IV} + 0.5u)}\right), \quad (20)$$

where $\alpha_{IV} = 1$ for Strong IV (S) and $\alpha_{IV} = 0.5$ for Weak IV (W). Unless otherwise noted, $\alpha_{IV}$ is set as 1. The outcome variable is generated with setting $(A)$ for Poisson and Negative Binomial and with setting $(B)$ in Table 1 for a Neyman Type A outcome. There were 21.6% zeros in the generated Poisson outcome, 25.6% zeros in the generated Negative Binomialoutcome and 52.7% zeros in the generated Neyman Type A distribution outcome. We considered a standard normal baseline covariate $X$ and the corresponding IV $Z \times X^{IV}$. We have eight estimating equations in (16) for four parameters. Two computational methods are proposed to obtain estimates for the parameters:

(a) EE-EL1: The first four estimating functions $g_1, ..., g_4$ in (16) are incorporated into the empirical likelihood for estimates and the next four estimating equations $g_5, ..., g_8$ are used to evaluate the goodness of fit of the estimates.

(b) EE-EL2: All eight estimating functions $g_1, ..., g_8$ in (16) are incorporated into the empirical likelihood for estimates.

The comparisons in the supplementary material Table 6 show that both EE-EL1 and EE-EL2 work well. EE-EL1 is fast in terms of computation while EE-EL2 has better performance in terms of median absolute deviation (MAD). Simulation studies reported in this paper were performed with computationally efficient EE-EL1 while real data analyses were conducted with the more stable EE-EL2. Table 2 shows the median and MAD of the es-timates from the Estimating Equations and Empirical Likelihood (EE-EL), Two Stage Residual Inclusion (2SRI), and ordinary regression (Reg) for direct and indirect effect parameters ($\beta_z$ and $\beta_m$). The ordinary regression fits a Poisson, Negative Binomial, and zero-inflated Poisson model respectively for the outcome on treatment, mediator and covariates. The 2SRI fits a logistic first stage model for the binary mediator with the IV $Z \times \mathbf{X}^{IV}$, and then fits a Poisson, Negative Binomial, and zero-inflated Poisson model for the outcome. The EE-EL can be generally used for Poisson, Negative binomial outcome and Neyman Type A distribution without specifying the distribution. As shown in Table 2, the ordinary regression estimates (Reg) are generally biased while both 2SRI and EE-EL estimates have reduced bias with the use of the IV. When the mediator is binary, 2SRI estimate has small bias on the direct effect parameter $\beta_z$ for the Poisson and Negative Binomialmodels, but can have a bias of greater than 15% for the Neyman Type A outcome when there is a large proportion of zeroes. For the controlled indirect effect parameter $\beta_m$, 2SRI can have a bias of greater than 25% for the Poisson and Negative Binomialmodels and a bias of greater than 100% for the Neyman Type A model. The EE-EL estimator performed best for all the settings and the small bias diminished with increased sample size and stronger IV.

Single Continuous Mediator As we discussed in Section 2.5, the natural indirect effect ratio (9) can be estimated for a continuous mediator. The mediator is generated as $m = -0.5 + 0.5z + 0.5x + \alpha_3 z \times x^{IV} + 0.5u + v$, where $v$ follows a standard normal distribution and the outcome variable is generated with setting $(A)$ for Poisson and Negative Binomial and with setting $(B)$ in Table 1 for Neyman Type A outcome. Table 3 shows the median estimates and MAD for natural direct and indirect effect ratios with a continuous mediator. The ordinary regression estimates are biased while both the 2SRI and EE-EL estimates are approximately unbiased. The results for controlled direct and indirect effect ratios with a continuous mediator (see the supplementary material Table 2) are similar to the results with a binary mediator in Table 2.

Multiple Instrumental Variables We also considered settings with more than one instrumental variable. For example,

$$m|x,z,u \sim Ber\left(\frac{\exp(-0.5+0.5z+0.5x_1+0.5x_2+z\times x_1^{IV}+z\times x_2^{IV}+0.5u)}{1+\exp(-0.5+0.5z+0.5x_1+0.5x_2+z\times x_1^{IV}+z\times x_2^{IV}+0.5u)}\right). \quad (21)$$

The results (see the supplementary material Table 1) are similar to Table 2. The EE-EL estimates are consistent while the 2SRI estimates have large bias in some settings.

Multiple Binary Mediators Considering settings with two independent mediators and two IVs $Z \times X_1^{IV}$ and $Z \times X_2^{IV}$, mediators $m_1$ and $m_2$ are generated independently as

$$m_1|x,z,u \sim Ber\left(\frac{\exp(-0.5+0.5z+0.5x_1+0.5x_2+z\times x_1^{IV}+z\times x_2^{IV}+0.5u)}{1+\exp(-0.5+0.5z+0.5x_1+0.5x_2+z\times x_1^{IV}+z\times x_2^{IV}+0.5u)}\right), \quad (22)$$

$$m_2|x,z,u \sim Ber\left(\frac{\exp(1+z-0.5x_1+x_2+z\times x_1^{IV}+z\times x_2^{IV}+0.5u)}{1+\exp(1+z-0.5x_1+x_2+z\times x_1^{IV}+z\times x_2^{IV}+0.5u)}\right). \quad (23)$$

The outcome variable is generated as

$$\mathbb{E}\left(Y(z,m_1,m_2)|z,m_1,m_2,x,u\right) = \exp\left(\beta_0+\beta_z z+\beta_{m,1}m_1+\beta_{m,2}m_2+\beta_x x+\beta_u u\right), \quad (24)$$

with setting $(C)$ for Poisson and Negative Binomial and with setting $(D)$ in Table 1 for Neyman Type A outcome. There were 26.3% zeros in the generated Poisson outcome, 29.8% zeros in the generated Negative Binomial outcome and 38.6% zeros in the generated Neyman Type A distribution outcome. Similar to Table 2, Table 4 shows that the ordinary regression estimate is heavily biased. 2SRI can reduce the bias in some cases but has large bias in other cases when the sample size is small and/or the percentage of zeros is relatively large. The EE-EL performed best in all the cases and produced consistent estimates with increased sample size.

Sensitivity Analysis Method. Sensitivity analysis was examined as discussed in Section 3.3. The sensitivity parameter $\eta$ represents how much the proposed instruments violate the assumptions needed for them to be valid instruments. The supplementary material Table 3 shows that after we adjust the outcome and use the adjusted outcome in the methods, the results are similar to the results in Table 1 and the EE-EL estimates are approximately unbiased for the direct and indirect effect parameters. Thus, if the amount of violation of the assumptions is correctly specified by $\eta$, the sensitivity analysis method provides unbiased estimates of the direct and indirect effect parameters.

High Proportion of Zeros and Mis-specification in Outcome Distribution. Additional simulation studies were conducted to examine the performance of methods with increased percentage of zeros (50% for Poisson and 55% for Negative Binomial), and when the outcome distribution is misspecified in 2SRI. The results with increased percentage of zeros (the supplementary material Table 4) are similar to the results in Table 2. When the outcome distribution is misspecified, 2SRI produced biased estimates while EE-EL continues to perform well as it does not rely on a parametric outcome distribution (the supplementary material Table 5).

In summary, the ordinary regression analysis produces biased estimates for the direct and indirect effect parameters when there is unmeasured confounding. The 2SRI and EE-EL reduce bias with the use of instrumental variables. However, 2SRI can have a large bias when the sample size is small or the percentage of zeros is large with a binary mediator or the outcome distribution is misspecified. The EE-EL method generally performs well under different settings and is robust to the misspecification of outcome distribution. The sensitivity analysis we proposed performs well too.

## 5. Real Data Analysis

We analyzed two real studies with methods developed in this paper, one study with a binary mediator and another study with a continuous mediator.

Dental Study DDHP MI-DVD. In the Detroit Dental Health Project's Motivational Interviewing DVD (DDHP MI-DVD) trial (Ismail et al. 2011), 790 families (0-5 years old children and their caregivers) were randomly assigned to one of two education groups (DVD only or MI+DVD). Both groups of families received a copy of a special 15-minute DVD for dental education. Additionally the families in the intervention group (MI+DVD) met an motivational interviewing (MI) interviewer, developed their own preventive goals, and received booster calls within 6 months of the intervention. The primary analyses of the study showed that caregivers in the MI+DVD group were more likely to make sure their child brushed at bedtime at 6 months and 2 years, but the intervention did not have a significant effect on children's dental outcomes at 2 years. In this study, we are interested in whether there was a direct effect of the intervention on children's dental outcomes that cancelled out a mediation effect in the opposite direction so that no significant total effect of the intervention was found. We assessed if the intervention had an effect on the outcome (the number of new untreated lesions at 2 years) through a binary mediator (whether or not caregivers made sure their child brushed at bedtime). The outcome has a large proportion of zeros – more than 60% of children had zero new untreated lesions (Figure 1(a)). The proposed instrumental variables we consider are the interactions between intervention and three baseline covariates: number of times child brushed, whether or not caregivers made sure their child brushed at baseline, and whether or not caregivers provided the child healthy meals at baseline. A logistic regression was fit for the binary mediator, showing significant effect of both the treatment and IV.

Although we are not able to test all of the ways in which the ER as-sumption of the IVs might be violated, we examined the plausibility of the IVs satisfying the ER assumption by examining whether the IV had effects through other pathways, which would suggest that the IVs violate the ER assumption; see Section 3.3 and the supplementary material Section 1.5. Specifically we examined whether the IVs were associated with other intermediate variables such as whether the caregiver provided the child with non-sugared snacks, whether the caregiver gave the child healthy meals, whether the caregiver checked the child for early non-cavitated demineralized enamel, whether the caregiver made sure the child saw a dentist every 6 months given the intervention and baseline covariates. None of the IVs were significantly associated with other intermediate variables, indicating no evidence of the violation of the ER assumption of instrumental variables.

Table 5 summarizes the EE-EL estimates of the direct and indirect effect ratios and the bootstrap confidence intervals. Note that a ratio of 1 indicates no effect. The result shows that the intervention did not have much direct effect on the number of new untreated lesions (controlled direct effect ratio 1.081), and parent behavior in making sure their child brushed at bedtime tended to decrease the number of new untreated lesions (controlled indirect effect ratio 0.595) but the effect was not statistically significant with a 90% CI $(0.0524, 9.735)$.

In this paper we are particularly interested in whether or not the MI intervention affected the children's oral health through its effect on the mediator whether or not caregivers made sure their child brushed at bedtime. Besides the pathway through the mediator, the effects of MI intervention on the children's oral health through other pathways are included in the direct effect.

The sensitivity analysis (Figure 2) shows that with increased amount of violation of the ER assumption, that is, with increased direct effect of the IV ($Z \times X^{\mathrm{IV}}$) on children's oral health ($\eta$ increases from 0 to 0.15), the direct and indirect effects of the MI intervention on children's oral health decreased. Specifically, the direct effect changed from a small increase to a small reduction in the number of new untreated lesions (the direct effect ratio drops below 1 from 1.081 but neither was significant), and indirect effect through parents making sure their chid brushed at bedtime showed additional reduction in the number of untreated lesions (the controlled indirect effect ratio decreases from 0.595 to 0.147).

Flood data analysis. In 1998, two-thirds of Bangladesh suffered from massive floods. del Ninno et al. (2001) conducted a study of the effects of flooding on health outcomes. We will use our method to see whether a household being severely affected by the flood (treatment) influenced the number of days a child had diarrhea in the three month period after the flood (outcome) through its effect on the per capita calorie consumption of the household (mediator). The outcome histogram Figure 1(b) shows that more than 70% children had zero days of diarrhea.

We assume strong ignorability for the treatment conditioning on baseline covariates: sex, age, the size of the household, mother's education, father's education, indicator of missing values for mother's education and father's education, mother's age, father's age, indicator of missing values for mother's age and father's age. Because the mediator is continuous, we are able to evaluate the natural effects as discussed in Section 2.5. The proposed instrumental variable we consider is interaction of the flooding and a baseline covariate, whether the household has a low or large amount of farmland available. The treatment and the instrumental variable have significant effects on the mediator.

We partially tested whether the proposed instrumental variable satisfies the exclusion restriction as described in Section 3.3. Specifically, in addition to the flood effect on the children's number of days of diarrhea through its effect on the per capita calorie consumption of the household, the flood could also affect children's number of days of diarrhea through other pathways, which were not the interest in this paper and therefore included in the direct effect. One possible pathway is mother's health. del Ninno et al. (2001) used the variable, whether a mother had chronic energy deficiency (CED) as a measure of mother's health. Specifically a mother was classified as being CED if her body mass index was less than 18.5. Non-significant association of the IV with CED (p=0.1501) given flood and baseline covariates indicated no evidence of violation of the ER assumption.

Table 5 illustrates the EE-EL estimates of the controlled (natural) direct and indirect effect and the bootstrap confidence intervals. The flooding tended to increase but did not have a significant direct effect on the number of days of diarrhea (direct effect ratio 1.229, 90% CI: 0.300, 2.409). A larger per capita calorie consumption of a household led to a significant decrease in the number of days a child had diarrhea over the three month period after the flood (controlled indirect effect ratio 0.040, 90% CI: $2.524 \times 10^{-5}$, 0.822); and decreased per capita calorie consumption of a household due to flooding led to a significant increase in the number of days a child had diarrhea after the flood (natural indirect effect ratio 1.685, 90% CI: 1.010, 6.239).

We further conducted a sensitivity analysis to see how the results would change with the violation of the ER assumption. Figure 2 shows that with increased direct effect of $Z \times \mathbf{X}^{\mathrm{IV}}$ on children's diarrhea, the direct effect of flood changed from a small increase 1.229 to a small reduction 0.934 in the number of days of diarrhea with neither significant, and the indirect effect through higher per capita calorie consumption showed a greater reduction in the number of days of diarrhea (from 0.040 to 0.013). With increased amount of the violation of the ER assumption, the increased indirect effect indicates a greater rise in the number of days of diarrhea from decreased per capita calorie consumption of a household due to flooding.

## 6. Conclusion and discussion

In this paper, we consider mediation analysis and define the direct and indirect effect ratios for a count or zero inflated count outcome when there is concern about unmeasured confounding between the mediator and outcome. Our method uses the interaction of treatment and baseline covariates as instrumental variables, constructs estimating equations and use the empirical likelihood approach to combine the information in estimating equations. Our method relaxes the assumption of sequential ignorability with reasonable assumptions and does not rely on parametric outcome distribution assumptions. A sensitivity analysis is proposed for the violation of the ER assumption. Simulation studies show that the two stage approach (2SRI) reduces bias with the use of IV compared to ordinary regression, but can produce biased estimates when the mediator is binary. The estimating equations empirical likelihood (EE-EL) method generally provides approximately unbiased estimates for the direct and indirect effects for different types of outcomes and under different settings (binary mediator, continuous mediator, multiple independent mediators, multiple instrumental variables), and is robust to the outcome distribution.

The model (5) considered in this paper can be generalized to include the interaction terms $z \times \mathbf{x}$,$z \times m$ and $\mathbf{x} \times m$.

$$\begin{aligned} & f\{\mathbb{E}\left(Y(z,m)|\mathbf{x},u\right)\} \qquad (25) \\ = \quad & \beta_0 + \beta_z z + \beta_m m + \beta_{\mathbf{x}}^T \mathbf{x} + \beta_u u + \beta_{z\mathbf{x}}^T \mathbf{x} \times z + \beta_{\mathbf{x}m}^T \mathbf{x} \times m + \beta_{zm} z \times m, \end{aligned}$$

where $f$ is a link function and $\mathbf{x}$ is a random vector of length $l$. In model (26), we have $l+2$ endogenous variables $m$, $\mathbf{x}\times m$ and $z\times m$. Hence, identification requires at least $l+2$ valid instruments $z \times \mathbf{x}^{IV}$ such that conditioning on $z, \mathbf{x}$ and $z \times \mathbf{x}$, (1) $z \times \mathbf{x}^{IV}$ predict the endogenous variables $m$, $\mathbf{x} \times m$ and $z \times m$; (2) $z \times \mathbf{x}^{IV}$ is independent of the unmeasured confounding $u$; and

(3) $z \times \mathbf{x}^{IV}$ does not have a direct effect on the outcome other than through its effects on $m$,$\mathbf{x} \times m$ and $z \times m$. However, the effects are not the same as (9)-(11), but should be re-derived from the generalized model (26) and (1)-(4). This is beyond the scope of this paper and we will leave it for a further study.

**Supplementary Materials**

Web Appendices, Tables, and proofs are available at the Biometrics website on Wiley Online Library.

**Acknowledgement**

The authors thank Amid Ismail and Sungwoo Lim for providing us the DDHP MI-DVD data. This paper was made possible by Grant Number U54 DE 019285 from the NIDCR, a component of the National Institutes of Health (NIH).

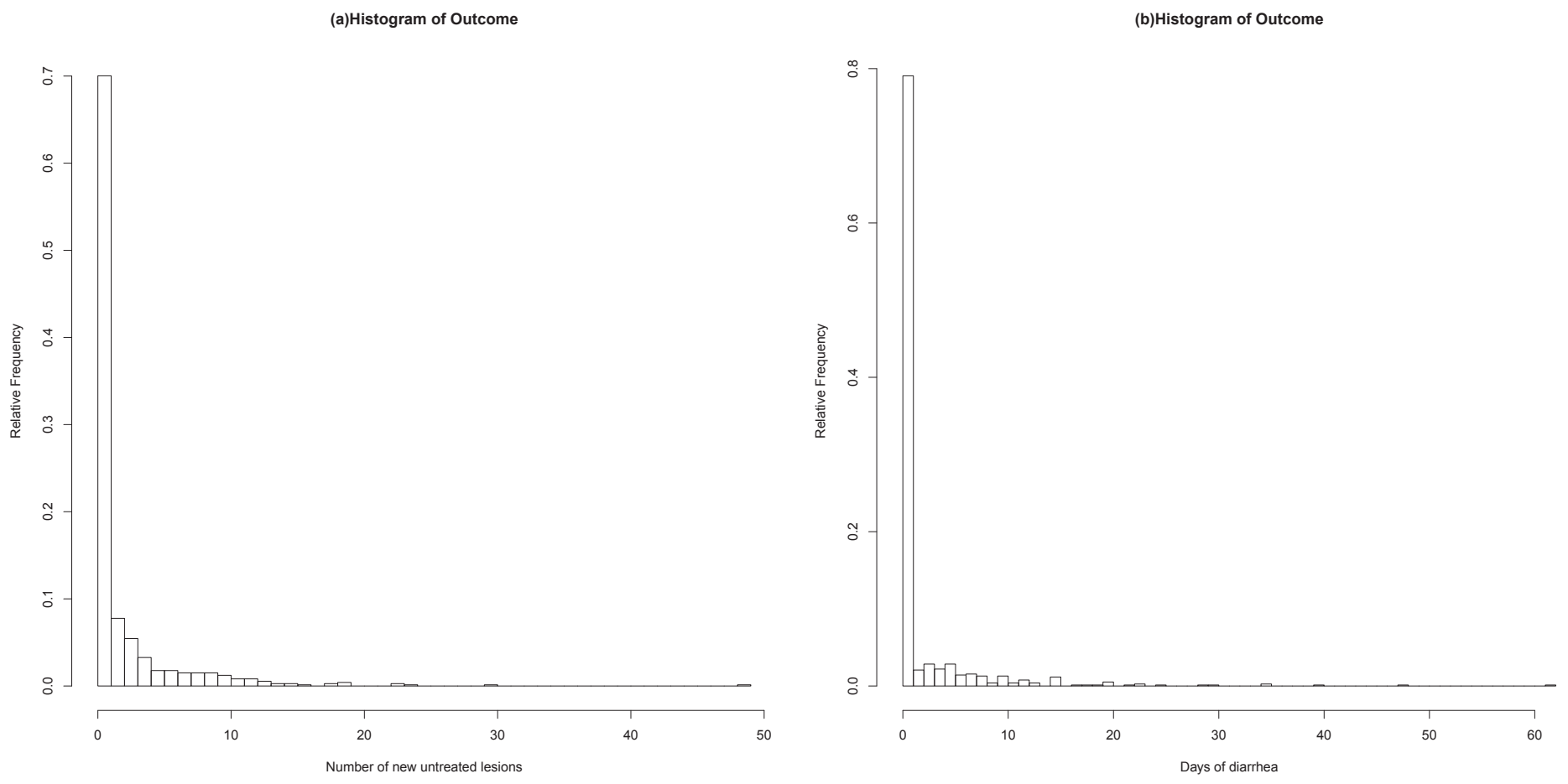

Fig. 1: (a) Dental Study; (b) Flood data;

| Setting | $\Theta$ | value | $\Theta$ | value | $\Theta$ | value | $\Theta$ | value | $\Theta$ | value |
|---|---|---|---|---|---|---|---|---|---|---|
| (A) | $\beta_0$ | 1 | $\beta_z$ | 0.5 | $\beta_m$ | 0.5 | $\beta_{\mathbf{x}}$ | 0.5 | $\beta_u$ | 1 |
| (B) | $\gamma_0$ | 0.6 | $\gamma_z$ | 0.25 | $\gamma_m$ | 0.25 | $\gamma_{\mathbf{x}}$ | 0.25 | $\gamma_u$ | 0.5 |
| | $\lambda_0$ | -1 | $\lambda_z$ | 0.25 | $\lambda_m$ | 0.25 | $\lambda_{\mathbf{x}}$ | 0.25 | $\lambda_u$ | 0.5 |
| (C) | $\beta_0$ | 1 | $\beta_z$ | 0.5 | $\beta_{m_1}$ | 1 | $\beta_{\mathbf{x}_1}$ | 0.5 | $\beta_u$ | 1 |
| | | | | | $\beta_{m_2}$ | 0.5 | $\beta_{\mathbf{x}_2}$ | 0.5 | | |
| (D) | $\gamma_0$ | 1 | $\gamma_z$ | 0.25 | $\gamma_{m_1}$ | 0.75 | $\gamma_{\mathbf{x}_1}$ | 0.5 | $\gamma_u$ | 0.5 |
| | | | | | $\gamma_{m_2}$ | 0.25 | $\gamma_{\mathbf{x}_2}$ | 0.5 | $\gamma_u$ | 0.5 |
| | $\lambda_0$ | -0.5 | $\lambda_z$ | 0.25 | $\lambda_{m_1}$ | 0.25 | $\lambda_{\mathbf{x}_1}$ | 0.5 | $\lambda_u$ | 0.5 |
| | | | | | $\lambda_{m_2}$ | 0.25 | $\lambda_{\mathbf{x}_2}$ | 0.5 | | |

Table 1: Parameter settings in simulation studies

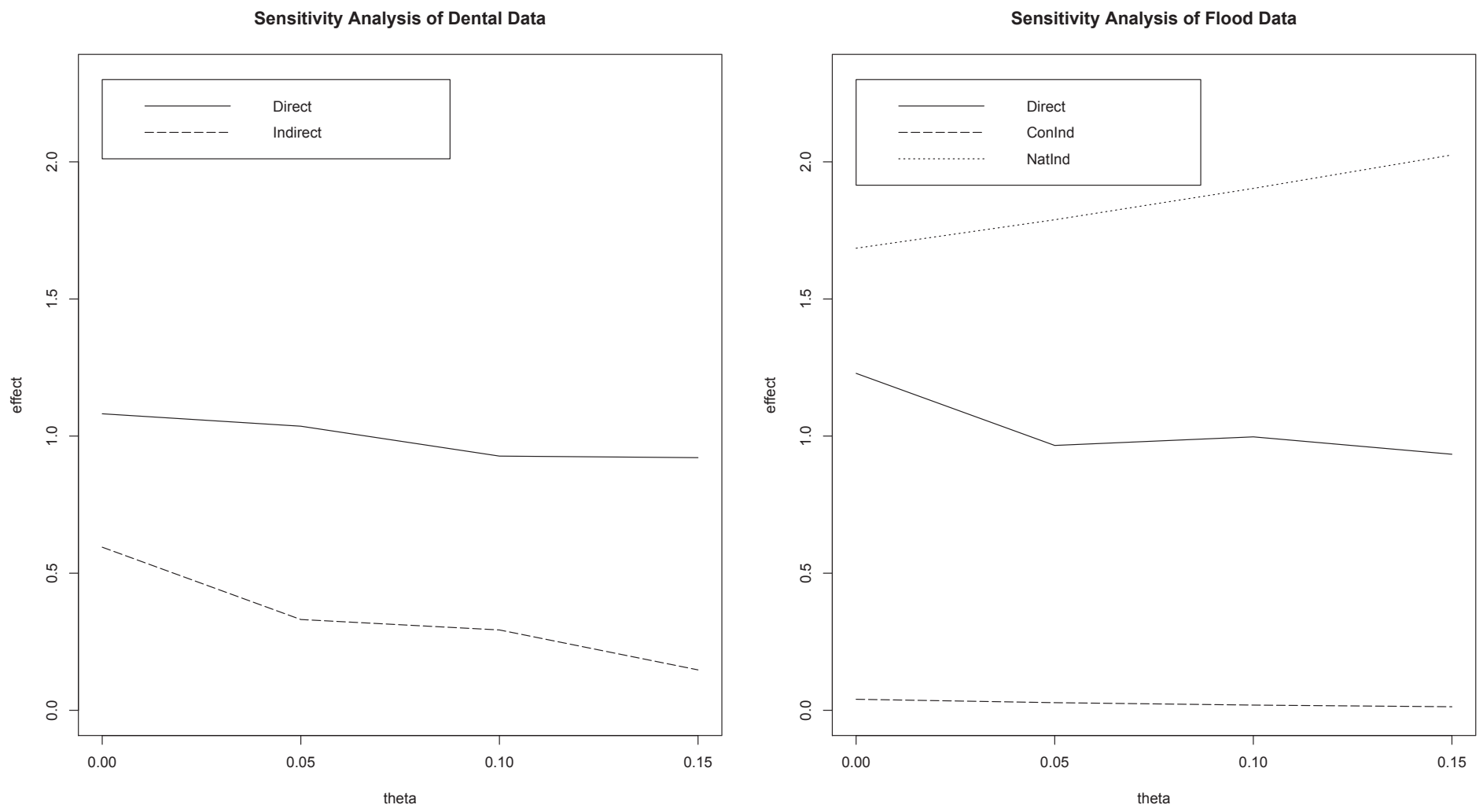

Fig. 2: Sensitivity analysis of the real data.
Left: "Direct" – the controlled and natural direct effect; "Indirect" – the controlled indirect effect.
Right: "Direct" – the controlled and natural direct effect; "ConInd" – the controlled indirect effect; and "NatInd" – the natural indirect effect.

| | | | Direct | | | Indirect | | |
|---|---|---|---|---|---|---|---|---|
| | | | EE-EL | 2SRI | Reg | EE-EL | 2SRI | Reg |
| Out | IV | n | Med. | Med. | Med. | Med. | Med. | Med. |
| | | | (MAD) | (MAD) | (MAD) | (MAD) | (MAD) | (MAD) |
| Poi | S | 500 | 0.496 | 0.505 | 0.416 | 0.522 | 0.559 | 0.955 |
| | | | (0.126) | (0.150) | (0.117) | (0.691) | (0.722) | (0.102) |
| Poi | S | 1000 | 0.497 | 0.497 | 0.419 | 0.522 | 0.561 | 0.948 |
| | | | (0.094) | (0.111) | (0.080) | (0.489) | (0.495) | (0.082) |
| Poi | S | 5000 | 0.497 | 0.500 | 0.415 | 0.510 | 0.528 | 0.948 |
| | | | (0.039) | (0.047) | (0.034) | (0.225) | (0.224) | (0.036) |
| Poi | W | 500 | 0.489 | 0.500 | 0.426 | 0.530 | 0.524 | 0.966 |
| | | | (0.159) | (0.198) | (0.122) | (1.117) | (1.266) | (0.104) |
| Poi | W | 1000 | 0.486 | 0.483 | 0.430 | 0.575 | 0.643 | 0.962 |
| | | | (0.120) | (0.133) | (0.078) | (0.793) | (0.852) | (0.082) |
| Poi | W | 5000 | 0.500 | 0.496 | 0.432 | 0.499 | 0.576 | 0.959 |
| | | | (0.054) | (0.058) | (0.035) | (0.389) | (0.388) | (0.036) |
| NB | S | 500 | 0.494 | 0.503 | 0.443 | 0.448 | 0.494 | 0.944 |
| | | | (0.164) | (0.155) | (0.127) | (0.820) | (0.736) | (0.131) |
| NB | S | 1000 | 0.499 | 0.503 | 0.447 | 0.555 | 0.528 | 0.942 |
| | | | (0.114) | (0.116) | (0.086) | (0.615) | (0.515) | (0.089) |
| NB | S | 5000 | 0.498 | 0.499 | 0.442 | 0.486 | 0.502 | 0.945 |
| | | | (0.050) | (0.051) | (0.037) | (0.262) | (0.225) | (0.041) |
| NB | W | 500 | 0.493 | 0.500 | 0.447 | 0.543 | 0.526 | 0.952 |
| | | | (0.177) | (0.194) | (0.123) | (1.261) | (1.295) | (0.136) |
| NB | W | 1000 | 0.489 | 0.493 | 0.439 | 0.525 | 0.515 | 0.955 |
| | | | (0.134) | (0.130) | (0.084) | (0.983) | (0.849) | (0.094) |
| NB | W | 5000 | 0.499 | 0.499 | 0.449 | 0.524 | 0.547 | 0.951 |
| | | | (0.068) | (0.060) | (0.039) | (0.468) | (0.395) | (0.038) |
| NTA | S | 500 | 0.465 | 0.413 | 0.386 | 0.589 | 0.841 | 0.988 |
| | | | (0.316) | (0.478) | (0.228) | (1.759) | (1.723) | (0.206) |
| NTA | S | 1000 | 0.504 | 0.451 | 0.361 | 0.448 | 0.692 | 0.983 |
| | | | (0.209) | (0.395) | (0.174) | (1.256) | (1.365) | (0.157) |
| NTA | S | 5000 | 0.499 | 0.461 | 0.369 | 0.500 | 0.674 | 0.981 |
| | | | (0.119) | (0.242) | (0.088) | (0.639) | (0.783) | (0.073) |
| NTA | W | 500 | 0.476 | 0.407 | 0.414 | 0.515 | 1.054 | 1.012 |
| | | | (0.293) | (0.581) | (0.239) | (2.261) | (2.896) | (0.211) |
| NTA | W | 1000 | 0.496 | 0.412 | 0.403 | 0.436 | 0.903 | 0.988 |
| | | | (0.248) | (0.484) | (0.176) | (1.848) | (2.458) | (0.154) |
| NTA | W | 5000 | 0.499 | 0.448 | 0.400 | 0.500 | 0.766 | 0.982 |
| | | | (0.160) | (0.302) | (0.084) | (1.053) | (1.317) | (0.072) |

Table 2: Estimates for the direct effect parameter ($\beta_z = 0.5$) and the indirect effect parameter ($\beta_m = 0.5$) with one instrumental variable and one binary mediator.

n: sample size; Med: median of 1000 Monte Carlo estimates; MAD: median absolute deviance; EE-EL: estimating equations and empirical likelihood; 2SRI: two stage residual inclusion; Reg: ordinary regression; Out: Outcome distribution; Poi: Poisson; NB: Negative Binomial; NTA: Neyman Type A distribution outcome; S: stronger IV (setting 1); W: relatively weaker IV

| Out | IV | n | True | (9) with $x = 1$ EE-EL Med. (MAD) | 2SRI Med. (MAD) | Reg Med. (MAD) | True | (9) with $x = -1$ EE-EL Med. (MAD) | 2SRI Med. (MAD) | Reg Med. (MAD) |
|---|---|---|---|---|---|---|---|---|---|---|
| Poi | S | 500 | 2.117 | 2.114 | 2.053 | 3.163 | 0.779 | 0.788 | 0.798 | 0.684 |
| | | | | (0.487) | (0.619) | (0.491) | | (0.082) | (0.098) | (0.081) |
| Poi | S | 1000 | 2.117 | 2.088 | 2.100 | 3.169 | 0.779 | 0.784 | 0.788 | 0.682 |
| | | | | (0.320) | (0.455) | (0.388) | | (0.064) | (0.072) | (0.059) |
| Poi | S | 5000 | 2.117 | 2.112 | 2.098 | 3.203 | 0.779 | 0.779 | 0.781 | 0.678 |
| | | | | (0.152) | (0.219) | (0.184) | | (0.030) | (0.035) | (0.028) |
| Poi | W | 500 | 1.649 | 1.644 | 1.635 | 2.247 | 1.000 | 0.998 | 0.996 | 1.001 |
| | | | | (0.442) | (0.558) | (0.317) | | (0.063) | (0.062) | (0.117) |
| Poi | W | 1000 | 1.649 | 1.635 | 1.644 | 2.223 | 1.000 | 1.001 | 1.000 | 1.003 |
| | | | | (0.305) | (0.417) | (0.214) | | (0.047) | (0.047) | (0.082) |
| Poi | W | 5000 | 1.649 | 1.642 | 1.639 | 2.227 | 1.000 | 1.001 | 1.001 | 1.002 |
| | | | | (0.150) | (0.193) | (0.101) | | (0.023) | (0.022) | (0.037) |
| NB | S | 500 | 2.117 | 2.128 | 2.114 | 3.428 | 0.779 | 0.787 | 0.790 | 0.667 |
| | | | | (0.534) | (0.513) | (0.493) | | (0.093) | (0.089) | (0.088) |
| NB | S | 1000 | 2.117 | 2.114 | 2.101 | 3.422 | 0.779 | 0.781 | 0.787 | 0.667 |
| | | | | (0.379) | (0.357) | (0.343) | | (0.636) | (0.558) | (0.095) |
| NB | S | 5000 | 2.117 | 2.115 | 2.096 | 3.421 | 0.779 | 0.779 | 0.781 | 0.664 |
| | | | | (0.166) | (0.163) | (0.154) | | (0.065) | (0.063) | (0.059) |
| NB | W | 500 | 1.649 | 1.661 | 1.650 | 2.376 | 1.000 | 0.996 | 0.996 | 0.996 |
| | | | | (0.564) | (0.537) | (0.323) | | (0.063) | (0.059) | (0.127) |
| NB | W | 1000 | 1.649 | 1.681 | 1.670 | 2.385 | 1.000 | 1.000 | 1.000 | 1.001 |
| | | | | (0.386) | (0.364) | (0.229) | | (0.051) | (0.050) | (0.095) |
| NB | W | 5000 | 1.649 | 1.646 | 1.638 | 2.365 | 1.000 | 0.999 | 0.999 | 0.999 |
| | | | | (0.156) | (0.158) | (0.103) | | (0.023) | (0.023) | (0.040) |
| NTA | S | 500 | 2.117 | 2.057 | 2.079 | 3.225 | 0.779 | 0.788 | 0.796 | 0.684 |
| | | | | (1.256) | (1.064) | (0.787) | | (0.175) | (0.149) | (0.094) |
| NTA | S | 1000 | 2.117 | 2.141 | 2.075 | 3.247 | 0.779 | 0.778 | 0.785 | 0.677 |
| | | | | (1.006) | (0.951) | (0.591) | | (0.136) | (0.127) | (0.070) |
| NTA | S | 5000 | 2.117 | 2.136 | 2.082 | 3.252 | 0.779 | 0.776 | 0.783 | 0.674 |
| | | | | (0.448) | (0.528) | (0.354) | | (0.055) | (0.071) | (0.035) |
| NTA | W | 500 | 1.649 | 1.585 | 1.610 | 2.237 | 1.000 | 0.999 | 0.996 | 0.999 |
| | | | | ( 1.092) | (1.050) | (0.378) | | (0.084) | (0.069) | (0.119) |
| NTA | W | 1000 | 1.649 | 1.654 | 1.609 | 2.225 | 1.000 | 0.998 | 0.997 | 0.998 |
| | | | | (0.889 | (0.799) | (0.280) | | (0.050) | 0.042) | (0.082) |
| NTA | W | 5000 | 1.649 | 1.627 | 1.631 | 2.233 | 1.000 | 1.000 | 1.000 | 1.000 |
| | | | | (0.455) | (0.522) | (0.155) | | (0.020) | (0.019) | (0.037) |

Table 3: Estimates for the natural indirect rate ratio (9) with one instrumental variable and one continuous mediator.

| | Direct | | | Indirect1 | | | Indirect2 | | |
|---|---|---|---|---|---|---|---|---|---|
| | EE-EL | 2SRI | Reg | EE-EL | 2SRI | Reg | EE-EL | 2SRI | Reg |
| Out | Med. | Med. | Med. | Med. | Med. | Med. | Med. | Med. | Med. |
| (n) | (MAD) | (MAD) | (MAD) | (MAD) | (MAD) | (MAD) | (MAD) | (MAD) | (MAD) |
| Poi | 0.448 | 0.464 | 0.319 | 0.932 | 1.156 | 1.425 | 0.661 | 0.644 | 0.911 |
| (500) | (0.218) | (0.275) | (0.127) | (0.911) | (0.827) | (0.122) | (1.127) | (0.866) | (0.155) |
| Poi | 0.482 | 0.489 | 0.319 | 0.910 | 1.094 | 1.417 | 0.527 | 0.623 | 0.897 |
| (1000) | (0.162) | (0.174) | (0.089) | (0.703) | (0.636) | (0.086) | (0.909) | (0.687) | (0.111) |
| Poi | 0.490 | 0.502 | 0.321 | 0.990 | 1.080 | 1.416 | 0.521 | 0.511 | 0.902 |
| (5000) | (0.077) | (0.085) | (0.040) | (0.388) | (0.298) | (0.038) | (0.630) | (0.294) | (0.050) |
| NB | 0.442 | 0.481 | 0.342 | 0.979 | 1.184 | 1.432 | 0.592 | 0.514 | 0.950 |
| (500) | (0.248) | (0.234) | (0.126) | (0.964) | (0.885) | (0.139) | (1.185) | (0.911) | (0.160) |
| NB | 0.470 | 0.504 | 0.340 | 0.910 | 1.076 | 1.431 | 0.579 | 0.441 | 0.939 |
| (1000) | (0.183) | (0.153) | (0.088) | (0.747) | (0.585) | (0.104) | (0.978) | (0.636) | (0.120) |
| NB | 0.492 | 0.508 | 0.340 | 0.972 | 1.088 | 1.427 | 0.528 | 0.437 | 0.944 |
| (5000) | (0.086) | (0.065) | (0.040) | (0.454) | (0.293) | (0.041) | (0.697) | (0.305) | (0.052) |
| NTA | 0.423 | 0.377 | 0.315 | 0.925 | 1.247 | 1.445 | 0.653 | 0.838 | 0.949 |
| (500) | (0.308) | (0.495) | (0.191) | (1.214) | (1.362) | (0.189) | (1.359) | (1.411) | (0.240) |
| NTA | 0.473 | 0.440 | 0.316 | 0.915 | 1.185 | 1.440 | 0.546 | 0.616 | 0.939 |
| (1000) | (0.257) | (0.357) | (0.132) | (0.974) | (1.041) | (0.126) | (1.181) | (0.909) | (0.165) |
| NTA | 0.490 | 0.480 | 0.311 | 0.970 | 1.099 | 1.436 | 0.492 | 0.552 | 0.917 |
| (5000) | (0.119) | (0.201) | (0.066) | (0.541) | (0.560) | (0.059) | (0.801) | (0.464) | (0.077) |

Table 4: Estimates for the direct effect parameter ($\beta_z = 0.5$) and the indirect effect parameter ($\beta_{m_1} = 1$ and $\beta_{m_2} = 0.5$ ) with two IVs and two binary mediators.

| Dental study data | | |
|---|---|---|
| | EE-EL | 90% confidence interval |
| Controlled/Natural Direct | 1.081 | (0.760, 1.442) |
| Controlled Indirect | 0.595 | (0.052, 9.735) |
| **Flood data** | | |
| | EE-EL | 90% confidence interval |
| Controlled/Natural Direct | 1.229 | (0.299, 2.409) |
| Controlled Indirect | 0.040 | ($2.524 \times 10^{-5}$, 0.822) |
| Natural Indirect | 1.685 | (1.010, 6.239) |

Table 5: EL estimate and Bootstrap confidence interval for direct effect rate ratio and inderict effect rate ratio.